\documentclass[twocolumn]{aastex62}
\usepackage{color}

\graphicspath{{./}{figures/}}

\received{***}
\revised{***}
\accepted{***}
\submitjournal{AJ}

\shorttitle{Mid-IR observations of comet 17P/Holmes immediately after its great outburst}
\shortauthors{Shinnaka et al.}

\begin{document}

\title{Mid-infrared spectroscopic observations of comet 17P/Holmes immediately after its great outburst in October 2007}
\shortauthors{Shinnaka et al.}

\correspondingauthor{Yoshiharu Shinnaka}
\email{yoshiharu.shinnaka@cc.kyoto-su.ac.jp, yoshiharu.shinnaka@nao.ac.jp}

\author[0000-0003-4490-9307]{Yoshiharu Shinnaka}
\affil{Laboratory of Infrared High-resolution Spectroscopy (LiH), Koyama Astronomical Observatory, Kyoto Sangyo University, Motoyama, Kamigamo, Kita-ku, Kyoto 603-8555, Japan}
\affil{National Astronomical Observatory of Japan, 2-21-1 Osawa, Miataka, Tokyo 181-8588, Japan}

\author[0000-0002-5413-3680]{Takafumi Ootsubo}
\affiliation{Institute of Space and Astronautical Science (ISAS), Japan Aerospace Exploration Agency (JAXA), 3-1-1, Yoshinodai, Chuo-ku, Sagamihara, Kanagawa, 252-5210, Japan }

\author{Hideyo Kawakita}
\affiliation{Laboratory of Infrared High-resolution Spectroscopy (LiH), Koyama Astronomical Observatory, Kyoto Sangyo University, Motoyama, Kamigamo, Kita-ku, Kyoto 603-8555, Japan}
\affiliation{Department of Physics, Faculty of Science, Kyoto Sangyo University, Motoyama, Kamigamo, Kita-ku, Kyoto 603-8555, Japan}

\author{Mitsuru Yamaguchi}
\affiliation{Department of Physics, Faculty of Science, Kyoto Sangyo University, Motoyama, Kamigamo, Kita-ku, Kyoto 603-8555, Japan}

\author[0000-0002-6172-9124]{Mitsuhiko Honda}
\affiliation{Department of Physics, Kurume University School of Medicine, 67 Asahi-machi, Kurume, Fukuoka 830-0011, Japan}

\author{Jun-ichi Watanabe}
\affiliation{National Astronomical Observatory of Japan, 2-21-1 Osawa, Miataka, Tokyo 181-8588, Japan}



\begin{abstract}

Dust grains of crystalline silicate, which are rarely presented in interstellar space, were found in cometary nuclei. These crystalline silicates are thought to have formed by annealing of amorphous silicate grains or direct condensation of gaseous materials near the Sun in the solar nebula, and incorporated into cometary nuclei in the cold comet-forming region after radial transportation of grains in the solar nebula. Abundances of the crystalline silicate dust grains were therefore expected to be smaller farther from the Sun. We aim to better understand the formation mechanism of minerals incorporated into comet 17P/Holmes based on its mineral abundances. To derive the mineral composition of comet 17P/Holmes, we applied a thermal emission model for cometary dust grains to mid-infrared spectra of comet 17P/Holmes taken with the Cooled Mid-Infrared Camera and Spectrograph (COMICS) mounted on the Subaru Telescope a few days later the great outburst in October 2007. The resulting mass fraction of crystalline silicate, $f_{\rm cry}$, and an olivine-to-pyroxene abundance ratio, $f_{\rm OP}$, are $f_{\rm cry}$ = 0.31 $\pm$ 0.03 and $f_{\rm OP}$ = 1.20$^{+0.16}$/$_{-0.12}$, respectively. Based on a simple consideration of the mixing of dust grains originating in both the interstellar medium and solar nebula, the minerals of 17P/Holmes formed by non-equilibrium condensation. This result is consistent with theoretical and experimental predictions for vaporization and condensation of olivine in the solar nebula. 

\end{abstract}



\section{Introduction} \label{sec:intro}

The 10-$\mu$m emission feature is usually recognized in mid-infrared spectra of cometary dust coma when the comets are at around 0.1-3 au from the Sun \citep{HaBr04}, and the feature could be attributed to sub-$\mu$m-sized silicate grains \citep{Ma70}. Usually this spectral feature is composed of the broad features originating in amorphous silicates and the sub-peaks originating in crystalline silicates (\citealt{Ha94, Cr97, Wo99, Wo04, Ho04, Oo07a}, and references therein). Because there is no clear spectral evidence for crystalline silicates in interstellar medium \citep{Ke04}, the formation location for the crystalline silicate is still under debate. A possibility is that the crystalline silicates in comets were probably produced in the solar nebula from the interstellar amorphous silicate grains, by the process of thermal annealing or condensation in the inner region of the solar nebula (e.g., \citealt{Ha00, Fa00, HaDe02, Ga04}). Other possible mechanisms to make crystalline silicate in the space were proposed: an in situ annealing at a larger distance from the central star by a shock wave \citep{DeCo02, HaDe02}, an annealing inside the clump fragmented from a massive disk \citep{Vo11}, an episodic heating by intense radiation from a central star in outburst phase in surface layer of protoplanetary disks \citep{Ab09, Ju12}. These crystalline silicate grains were incorporated into cometary nuclei that formed far from the proto-Sun ($\sim$10-30 au) after the grains were transported to the outside of the solar nebula due to some mechanisms; e.g., the radial and vertical mixing of materials \citep{Boc02} or the X-wind \citep{Br12}. Therefore, the mass fraction of crystalline silicates with respect to the total (amorphous + crystalline) silicates is expected to be smaller for further distances from the Sun in the solar nebula \citep{Ga01, Boc02}. If the crystalline silicates formed by the direct condensation from gas-phase in the solar nebula \citep{Ga04, De17}, the forming region of the crystalline silicate grains was closer to the Sun than the case of thermal annealing process since the condensation of crystalline silicate grain requires higher temperatures ($\sim$1200-1400 K; \citealt{Ga04}) than the case of thermal annealing process for crystallization ($\sim$800 K; \citealt{Ga01}). In any case, the mass fraction of crystalline silicates in cometary grains is a clue to the place of comet-forming region in the solar nebula and the physical parameters of the solar nebula such as mass accretion rates and viscosity of the disk (e.g., \citealt{Ga01, Ga04, Boc02}). Namely, in addition to the mass fraction of crystalline silicate, abundance ratios of minerals composing cometary grains also give clues to the comet-forming region and the disk parameters. \citet{Ga04} theoretically demonstrated that abundance ratios of various minerals depend on distances from the proto-Sun. For the comet-forming region, olivine was estimated to be more abundant than pyroxene but the abundance ratios between olivine and pyroxene depend on distances from the proto-Sun according to their calculations. \citet{Li08} claimed that the abundance ratio between olivine to pyroxene silicate might be an indicator of system evolutionary age, based on mid-infrared spectroscopic observations of proto-planetary/circumstellar disks.

   Comet 17P/Holmes is classified as a Jupiter-family comet with an orbital period of 6.9 years. The comet was discovered by E. Holmes on UT 1892 November 6 \citep{Ho92}. The Tisserand criterion with respect to Jupiter is 2.86, computed from its latest orbital parameters listed in Nakano note\footnote{url{http://www.oaa.gr.jp/$\sim$oaacs/nk.htm}} (NK2484). The effective radius of the comet as spherical sphere was estimated as 1.71 km based on imaging photometric observations by the Hubble space telescope \citep{La00}. This value is similar to the effective radius of 1.62 $\pm$ 0.01 km (for the equivalent spherical body) with a large elongation of the nucleus with a large axial ratio of a/b $\geq$ 1.3 from the oscillations of brightness at large distances by using the 3.6-m New Technology Telescope \citep{Sn06}. The comet underwent a great outburst starting at UT 2007 October 23.3 \citep{Wa09, Hs10}, five months after perihelion at 2.05 au on UT 2007 May 5. This outburst reached maximum brightness of $\sim$2-3 mag in $V$-band as total magnitude within two days from the start of the outburst reported by many amateur astronomers. Such a huge outburst, which became brighter by 15 magnitudes, was unlike any other. Other than this outburst in 2007, comet 17P/Holmes had exhibited outbursts in November 1892 \citep{Ho92} and January 1893 \citep{Ba96}. 

According to \citet{Sc09}, the total amounts of water ice and dust grains during the 2007 apparition of 17P/Holmes were estimated to be $\sim$10$^{10}$ kg and 10$^{11}$ kg, respectively, corresponding to $\sim$0.2\% and at least 1-2\% of the total nucleus volume. \citet{Is10} reported the total mass injected into the coma by the outburst was estimated to be $>$4$\times$10$^{10}$ kg. \citet{Re10} found three different size components of dust grains and estimated total ejecta mass of $\sim$10$^{10}$ kg in the ejecta of the outburst in 2007, based on their mid-infrared spectroscopic and imaging observations by the $Spitzer$ Space Telescope. Note that their mid-infrared observations may slip past very small (sub-micron size) grains. \citep{Ya09} detected not only dust grains but also cold, icy, micron-size grains during the outburst. The enrichment of high-volatile gaseous species such as CH$_{3}$OH, C$_{2}$H$_{6}$, C$_{2}$H$_{2}$ in the coma was also demonstrated immediately after the outburst from the near-infrared spectroscopic observations \citep{DR08}. 

   Some studies have proposed mechanisms for those outbursts but are not yet conclusive; e.g., the phase-transition of amorphous water ice to crystalline water ice with a release of a huge amount of latent heat and gaseous volatiles \citep{Re10}, the vaporization of hypervolatile ice in the cometary nucleus \citep{Sc09}, the avalanche or landslide of comet surface \citep{Boi02, Br04}, and the POP model in which a cometary outburst was triggered by plugging pores and blocking the release of surface gas flow by the recrystallization of water in the surface regolith \citep{de16}. If a large amount of ice of amorphous water and/or hypervolatiles is required for the large-scale outbursts seen in comet 17P/Holmes, its nucleus was considered to form under lower temperature conditions (i.e., further distances from the Sun) than other normal comets in which no large-scale outburst was seen repeatedly. An enrichment of highly volatile species (such as CH$_{3}$OH, C$_{2}$H$_{6}$, C$_{2}$H$_{2}$) compared to other typical comet indicates that comet 17P/Holmes formed under lower-temperature conditions (further from the Sun in the solar nebula) and therefore it contains less crystalline silicate grains compared to other normal comets that did not show large-scale outbursts. To confirm this hypothesis, we derived the mass fraction of crystalline silicates in comet 17P/Holmes based on the mid-infrared low-resolution spectra of the comet taken immediately after its outburst in 2007.

\section{Observational data and data reduction} \label{sec:observation}

We reduced the imaging and low-resolution spectroscopic data of comet 17P/Holmes in the $N$-band (8-13 $\mu$m) wavelength obtained from the SMOKA archive system \citep{Ba02}. These data were taken by the Cooled Mid-Infrared Camera and Spectrograph (COMICS; \citealt{Ka00, Ok03, Sa03}) installed at the Cassegrain focus of the Subaru Telescope in Hawaii on UT 2007 October 25, 26, 28 and 28 immediately after its large-scale outburst in UT 2007 October 23 \citep{Wa09, Hs10}. Heliocentric and geocentric distances of the comet at the observations were 2.44-2.45 au and 1.63 au, respectively. A slit was used of length 40'' and width 0''.33 for the low-resolution spectroscopy mode of the $N$-band. This slit width corresponds to a spectral resolving power of $R \sim$250. The optical peak of the coma was placed on the slit for spectroscopic observations. The position angle of the slit length direction was set to 213$^\circ$ to observe both the nucleus center and the ejected dust cloud. To perform an absolute flux calibration and a correction of atmospheric absorption for the observed spectra of the comet, we also used imaging photometric data with the narrow-band filters $N$8.8 and $N$12.4 centered at 8.8 $\mu$m ($\Delta \lambda$ = 0.8 $\mu$m) and 12.4 $\mu$m ($\Delta \lambda$ = 1.2 $\mu$m), respectively, which are the standard filters installed in the COMICS. In addition to reducing the imaging and spectroscopic data of the comet, we reduced the photometric and spectroscopic data of HD 21552 as a standard star chosen from \citet{Co99}. The observed spectra of the standard star were compared with the template spectrum provided by \citet{Co99} for the absolute flux calibration. To cancel high-background radiation in N-band, the secondary-mirror-chopping was used at a frequency of 0.45-0.43 Hz with an amplitude of 60''. Directions of throw were 0$^\circ$, 240$^\circ$, 300$^\circ$ and 300$^\circ$ for UT 2007 October 25, 26, 27 and 28, respectively. Observational circumstances at our observations are listed in Table \ref{tab:table1}. Note that we should treat the data and results on UT 2007 October 27 carefully because thin clouds passed through the field-of-view during exposures.

\begin{deluxetable}{llccc}
\tablecaption{Observational circumstances of comet 17P/Holmes taken with the Subaru/COMICS \label{tab:table1}}
\tablehead{
\colhead{UT Time}  & \colhead{} & \colhead{$T_{\rm exp}$}  & \colhead{Observing mode} & \colhead{$r_{\rm H}$} \\
\colhead{}         & \colhead{} & \colhead{(s)}            & \colhead{}               & \colhead{(au)} }
\startdata
2007 Oct 25 & 10:12-10:44 & 430.1 & NLowSpec & 2.44 \\
            & 10:57       &  30.7 & $N$8.8   &      \\
            & 11:00       &  30.8 & $N$12.4  &      \\
2007 Oct 26 & 11:12-11:14 &  32.5 & $N$8.8   & 2.45 \\
            & 11:15-11:16 &  21.2 & $N$12.4  &      \\
            & 11:18-11:21 &  61.4 & NLowSpec &      \\            
2007 Oct 27 & 10:49-10:53 &  43.4 & $N$8.8   & 2.45 \\
            & 10:58-10:59 &  63.6 & $N$12.4  &      \\
            & 11:13-11:18 & 154.6 & NLowSpec &      \\
2007 Oct 28 & 10:19-10:28 &  75.9 & $N$8.8   & 2.45 \\
            & 10:30-10:34 &  63.6 & $N$12.4  &      \\
            & 10:58-11:13 & 245.6 & NLowSpec &      \\            
\enddata
\tablecomments{$T_{\rm exp}$ is total exposure time in seconds. $N$8.8 and $N$12.4 indicate the narrow-band filters centered at 8.8 $\mu$m ($\Delta\lambda$ of 0.8 $\mu$m) and 12.4 $\mu$m ($\Delta\lambda$ of 1.2 $\mu$m), respectively, for photometric observations. NLowSpec means a low-resolution spectroscopy in $N$-band with resolving power of $\sim$250. $r_{\rm H}$ is heliocentric distance at the observations in au. Imaging photometric data of $N$8.8 and $N$12.4 are used for calibrating an absolute flux of the comet.}
\end{deluxetable}

   The spectroscopic and photometric data in the mid-infrared wavelength region taken by the COMICS were reduced using IRAF software\footnote{IRAF is distributed by the National Optical Astronomy Observatory, which is operated by the Association of Universities for Research in Astronomy (AURA) under cooperative agreement with the National Science Foundation} as well as the special tools provided by the COMICS instrument team\footnote{\url{https://subarutelescope.org/Observing/DataReduction/index.html} and \url{
https://www.subarutelescope.org/Observing/Instruments/COMICS/canadia-mirror/comics/open/guide/index.html}}. We applied the standard reduction procedure (dark subtraction, chopping subtraction, and flat fielding) to the imaging data, performed the aperture photometry at the opt-center of the comet is performed to calibrate spectroscopic data (for the correction of different slit-loss in the observations of comet and standard star). As for the spectroscopic data, the standard chopping subtraction and flat-fielding by thermal spectra of the telescope cell cover and distortion correction were performed. To investigate the abundances of dust components of the comet signals, near nucleus regions were extracted for spectroscopic data. Extracted regions were $\pm$0''.83, $\pm$0''.91, $\pm$0''.91, and $\pm$0''.83 from the opt-center (considered as the near-nucleus region) on UT 2007 October 25, 26, 27, and 28, corresponding to $\pm$1,510 km, $\pm$1,660 km, $\pm$1,660 km, and $\pm$1,510 km from the opt-center, respectively. After combining all extracted spectra of the comet and the standard star, respectively, the wavelength was calibrated against atmospheric emission lines, and its uncertainty was estimated to be less than 0.02 $\mu$m. Finally, we scaled the flux of the comet spectra using the photometric data of the comet at the $N$8.8 and $N$12.4 bands calibrated by the standard star. An uncertainty of resultant spectra at each wavelength includes the Poisson noise caused by the high sky-background in the frames of the comet and the standard star, the readout noise of the detector estimated from blank regions of the frames, and a deviation of flux density of the template spectra provided by \citet{Co99} ($\sim$3\% in typical) as a random error.

\section{Thermal emission model of dust grains for comets} \label{sec:model}

We used the thermal emission model for cometary dust grains \citep{Oo07a} to derive physical properties of the grains. Following \citet{Ha02}, three types of minerals in amorphous phase (olivine, pyroxene, and carbon) and two types of minerals in crystalline phase (olivine and pyroxene) are considered as constituents. Our thermal emission model is similar to that of \citet{Ha02} but updated for some optical constants. The model parameters are optimized by minimizing $\chi^{2}$ between the observed and synthesized spectra in the wavelength range from 8.0 to 12.5 $\mu$m (see Figure \ref{fig:fig1}). A wavelength range from 9.4 to 9.8 $\mu$m is not used for the $\chi^{2}$-fitting to avoid the influence of strong telluric ozone (O$_{3}$) absorption band.

\begin{deluxetable*}{cccccccccc}
\tablenum{2}
\tablecaption{Fitting results of the thermal spectra of 17P/Holmes by using the thermal model\label{tab:table2}}
\tablewidth{0pt}
\tablehead{
\colhead{UT Date} & \multicolumn5c{Mass fraction of dust grains} & \multicolumn3c{Dust properties} & \colhead{$\chi^{2}_{\rm red}$}\\
\colhead{in 2007} & \multicolumn3c{Amorphous} & \multicolumn2c{Crystalline} & \colhead{$D$} & \colhead{$a_{\rm p}$ ($\mu$m)} & \colhead{$N$} & \colhead{}\\
\colhead{}        & \colhead{Olivine} & \colhead{Pyroxene} & \colhead{Carbon} & \colhead{Olivine} & \colhead{Pyroxene} & \colhead{}        & \colhead{}        & \colhead{}        
}
\startdata
Oct 25 & 0.31 $\pm$ 0.03           & 0.32 $\pm$ 0.03           & 0.12 $\pm$ 0.01 & 0.18 $\pm$ 0.01           & 0.08 $\pm$ 0.02           & 3.0 & 0.10$^{+0.01}$/$_{-0.00}$ & 3.34$^{+0.09}$/$_{-0.03}$ & 1.69 \\
Oct 26 & 0.37$^{+0.06}$/$_{-0.07}$ & 0.24$^{+0.05}$/$_{-0.04}$ & 0.13 $\pm$ 0.02 & 0.18 $\pm$ 0.03           & 0.08$^{+0.08}$/$_{-0.07}$ & 3.0 & 0.10$^{+0.03}$/$_{-0.00}$ & 3.07$^{+0.13}$/$_{-0.03}$ & 0.58 \\
Oct 27 & 0.26 $\pm$ 0.06           & 0.22 $\pm$ 0.05           & 0.14 $\pm$ 0.03 & 0.14$^{+0.03}$/$_{-0.02}$ & 0.23$^{+0.09}$/$_{-0.06}$ & 3.0 & 0.10$^{+0.12}$/$_{-0.00}$ & 2.93$^{+0.44}$/$_{-0.03}$ & 0.65 \\
Oct 28 & 0.39$^{+0.04}$/$_{-0.06}$ & 0.18 $\pm$ 0.03           & 0.09 $\pm$ 0.01 & 0.21 $\pm$ 0.02           & 0.13$^{+0.07}$/$_{-0.05}$ & 3.0 & 0.10$^{+0.04}$/$_{-0.00}$ & 3.00$^{+0.18}$/$_{-0.02}$ & 1.16 \\
Mean   & 0.32 $\pm$ 0.02           & 0.25 $\pm$ 0.02           & 0.11 $\pm$ 0.01 & 0.18 $\pm$ 0.01           & 0.09 $\pm$ 0.02 & --- & ---                       & ---                       & --- \\
\enddata
\tablecomments{Error ranges of each mass fraction are consistent with 68.3\% confidence level based on $\chi^{2}$ distribution in the degree of freedom of the thermal emission model of eight ($\Delta\chi^{2}$ = 9.31). Derived values of dust properties (e.g., $D$, $a_{\rm p}$, $N$) also mean best-fit and range of 68.3\% confidence level. Note that minimum size of dust grains in our fitting is 0.1 $\mu$m.}
\end{deluxetable*}

   In the thermal emission model for comet, we assume that all grains in the coma are at the same heliocentric distance (same as the heliocentric distance of the nucleus) and reach the temperature in radiative equilibrium condition with the solar radiation field. We consider the intrinsic properties for each grain (i.e., radius, composition, crystalline or amorphous form, porosity) and determine the temperature of each grain, by assuming an optically thin coma of cometary grain. An emission spectrum by a single grain is proportional to a product of emission efficiency factor ($Q_{\rm emit}$) and the Planck function ($B_\lambda (T_{\rm d})$; at an equilibrium temperature of the dust grain, $T_{\rm d}$). 
   To determine $T_{\rm d}$, we applied Mie theory with optical constants for each mineral: \citet{Do95} for amorphous olivine (MgFeSiO$_{4}$) and amorphous pyroxene (Mg$_{0.5}$Fe$_{0.5}$SiO$_{3}$), \citet{Pr93} for amorphous carbon (C), \citet{Fa01} for crystalline olivine (Mg$_{1.9}$Fe$_{0.1}$SiO$_{4}$), and \citet{Ja98} for crystalline pyroxene (MgSiO$_{3}$). 
As for calculation of the emission spectrum, the emission from crystalline silicates is difficult to reproduce using Mie theory \citep{YH99}. Then, to calculate the emission spectrum, in the case of crystalline silicates, we applied laboratory-measured mass absorption coefficients from \citet{Ko03} for crystalline olivine (Mg$_{2}$SiO$_{4}$) and \citet{Ch02} for crystalline pyroxene (MgSiO$_{3}$), although we applied  Mie theory with optical constants in the case of amorphous silicates (e.g., \citet{Do95} for amorphous olivine (MgFeSiO$_{4}$), amorphous pyroxene (Mg$_{0.5}$Fe$_{0.5}$SiO$_{3}$), and \citet{Pr93} for amorphous carbon (C)).
   An emission spectrum for each mineral is calculated as the sum of the fluxes integrated over a grain size distribution. Our model assumes the Hanner size distribution \citep{Ha83} as the grains size distribution. The Hanner size distribution is represented by $n(a) = (1-a_{0}/a)^{M} (a_{0}/a)^{N}$, where $a$ and $a_{0}$ are the grain radius and the minimum grain radius in $\mu$m. We assumed $a_{0}$ as 0.1 $\mu$m \citep{Oo07a, Ha02}. $N$ and $M$ are power-law indices of dust size distribution and related to the peak grain radius $a_{\rm p} = a_{0} (M+N)/N$. Thus, this form has two independent parameters: a peak grain radius ($a_{\rm p}$) and the slope of the size distribution of dust grains ($N$). For searching the best-fit parameters, we consider a grain size distribution with 0.1 $\mu$m $\leq a_{\rm p} \leq$ 0.5 $\mu$m in a step of 0.01 $\mu$m, 2.0 $\leq N \leq$ 4.0 in a step of 0.01, and porosity parameter ($D$) ranging from 2.5 to 3.0 ($D$ = 2.5, 2.609, 2.727, 2.857, and 3.0; \citealt{Oo07a}) corresponding to the fraction filled volume. 
   In the fitting, we assume that all mineral compositions have the same parameters of $a_{\rm p}$, $N$, and $D$, but amorphous and crystalline grains have radii in the range of 0.1-100 $\mu$m and of 0.1-1.0 $\mu$m, respectively, and mass fraction for each mineral is derived in the 0.1 - 1.0 $\mu$m range (same range as \citealt{Ha02} and \citealt{Wo11}). For simplicity, we also assume the spherical grains, although this assumption for the grain shape is probably not real because aggregates of smaller and elongated grains with variable sizes were found by in situ measurements of dust particles in 67P/Churyumov-Gerasimenko \citep{Be16}. A radiative equilibrium temperature of the crystalline olivine grains is 1.9 times higher than a computed equilibrium temperature in our model, according to the previous study of mid-infrared observations of comet C/1995 O1 (Hale-Bopp) \citep{Ha02}. Finally, a thermal spectrum of cometary dust particles in $N$-band, $F_{\rm obs}(\lambda)$ in W m$^{-2}$ $\mu$m$^{-1}$ as a function of wavelength $\lambda$ in $\mu$m, is expressed in both the thermal flux density per unit mass and mass of five dust grain components as follows: $F_{\rm obs}(\lambda) = m_{\rm AmoOl}F_{\rm AmoOl}(\lambda)+ m_{\rm AmoPy}F_{\rm AmoPy}(\lambda) + m_{\rm AmoCa}F_{\rm AmoCa}(\lambda) + m_{\rm CryOl}F_{\rm CryOli}(\lambda) + m_{\rm CryPy}F_{\rm CryPy}(\lambda)$, where $F_{\rm X}$ and $m_{\rm X}$ are flux densities per unit mass in W m$^{-2}$ $\mu$m$^{-1}$ kg$^{-1}$ and total mass in kg for each dust grain component, respectively. Subscript X means the five dust grain components: “AmoOl” for amorphous olivine, “AmoPy” for amorphous pyroxene, “AmoCa” for amorphous carbon, “CryOl” for crystalline olivine, or “CryPy” for crystalline pyroxene.   
   To convert unit of the thermal flux density to per unit mass, we use the density for each grain: 3.6$\times$10$^{3}$ kg m$^{-3}$ for amorphous olivine, 3.5$\times$10$^{3}$ kg m$^{-3}$ for amorphous pyroxene, 2.5$\times$10$^{3}$ kg m$^{-3}$ for amorphous carbon, 3.2$\times$10$^{3}$ kg m$^{-3}$ for crystalline olivine and crystalline pyroxene, respectively \citep{Li06}.
   There are eight free parameters considered in our thermal model: total mass of five kinds of dust grain components, $a_{\rm p}$, $N$, and $D$. Note that we cannot estimate absolute total masses of dust grains in the entire dust coma because our observation did not cover the comet's entire dust coma \citep{Is10} and derived values would have unknown systematic errors depending on the thermal emission model assumptions. Hereafter, we discuss relative mass fractions in the extracted inner coma region of 17P/Holmes.

\section{Results} \label{sec:results}

\begin{figure*}
\gridline{\fig{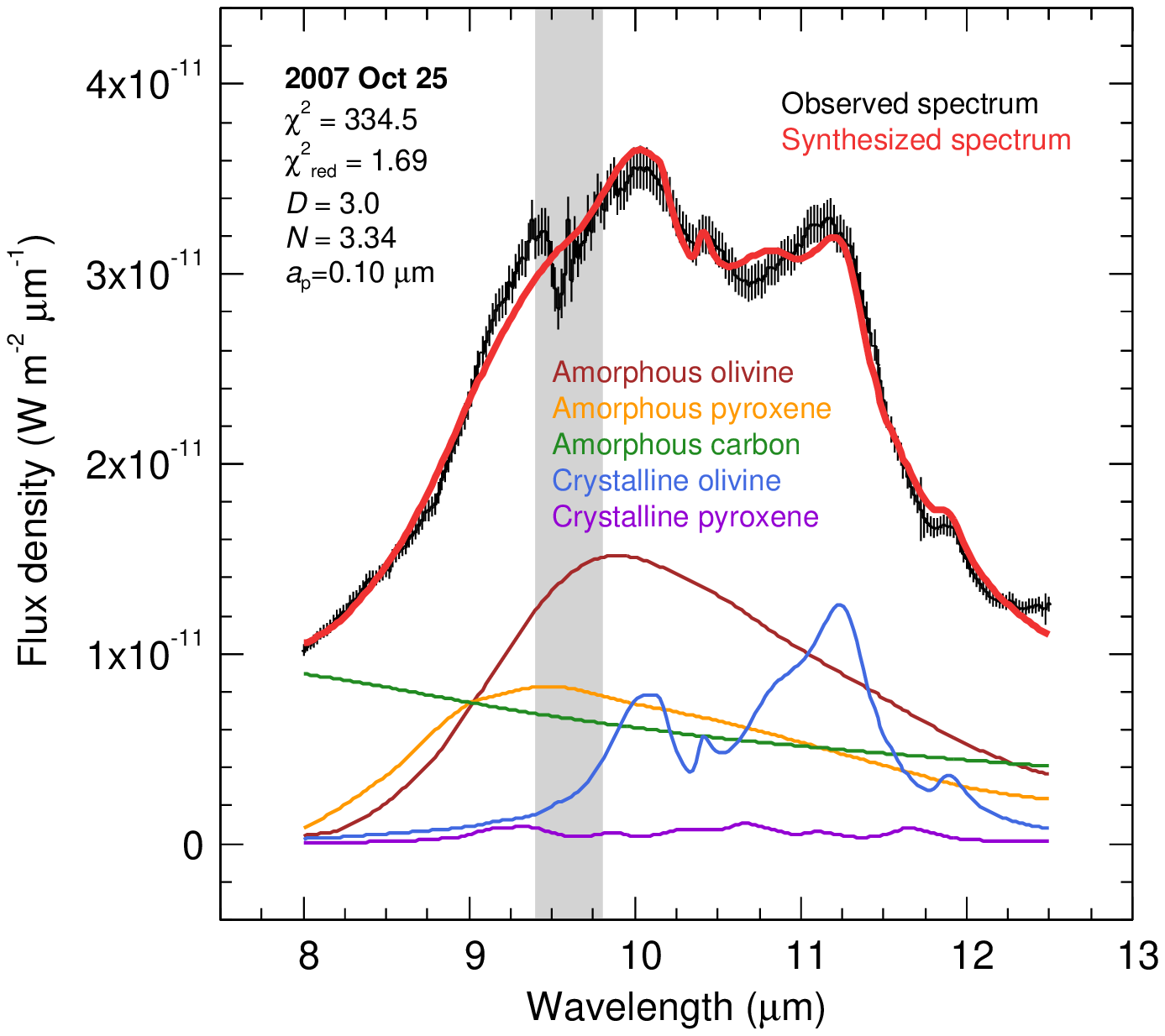}{0.45\textwidth}{(a)}
          \fig{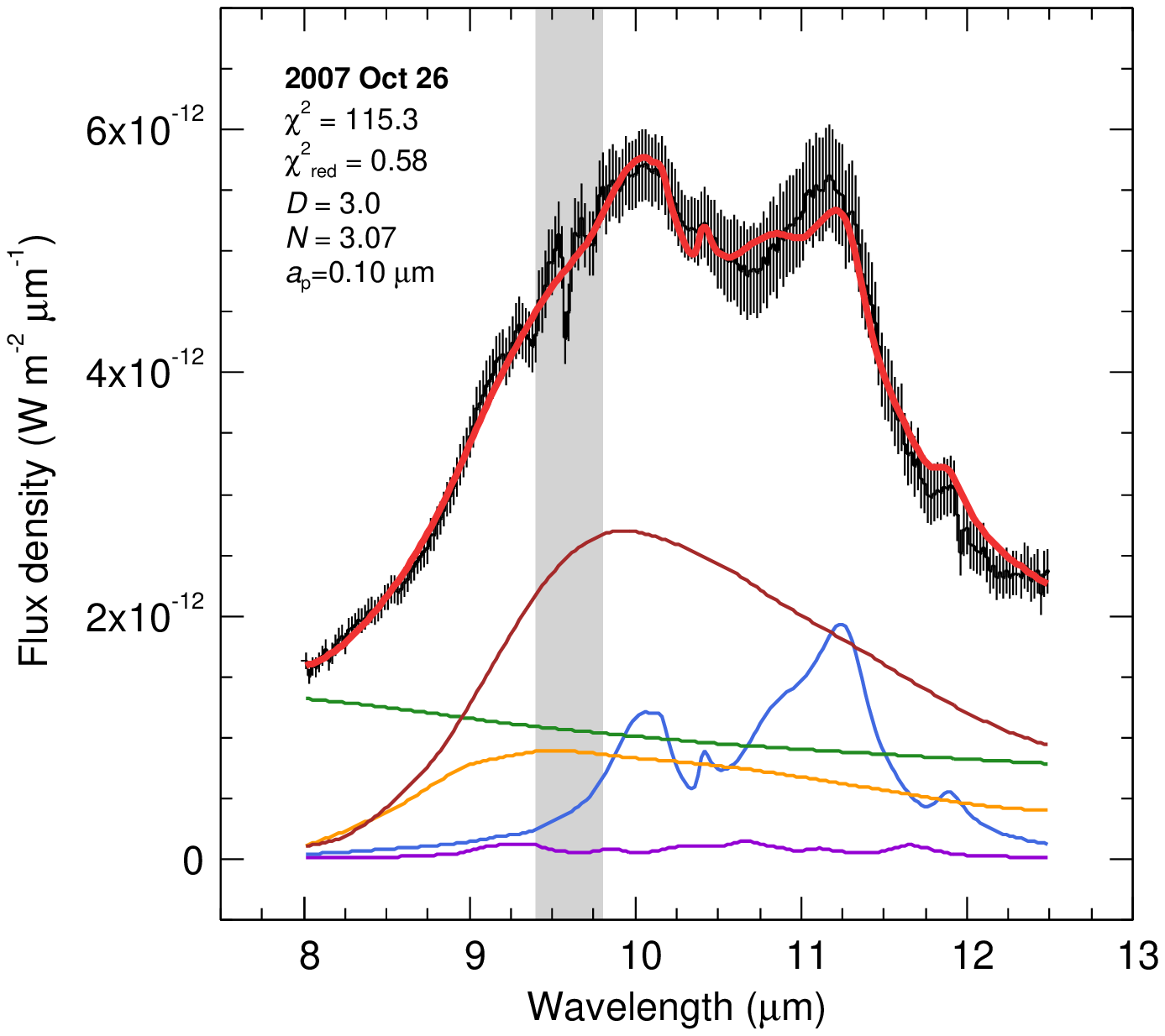}{0.45\textwidth}{(b)}
          }
\gridline{\fig{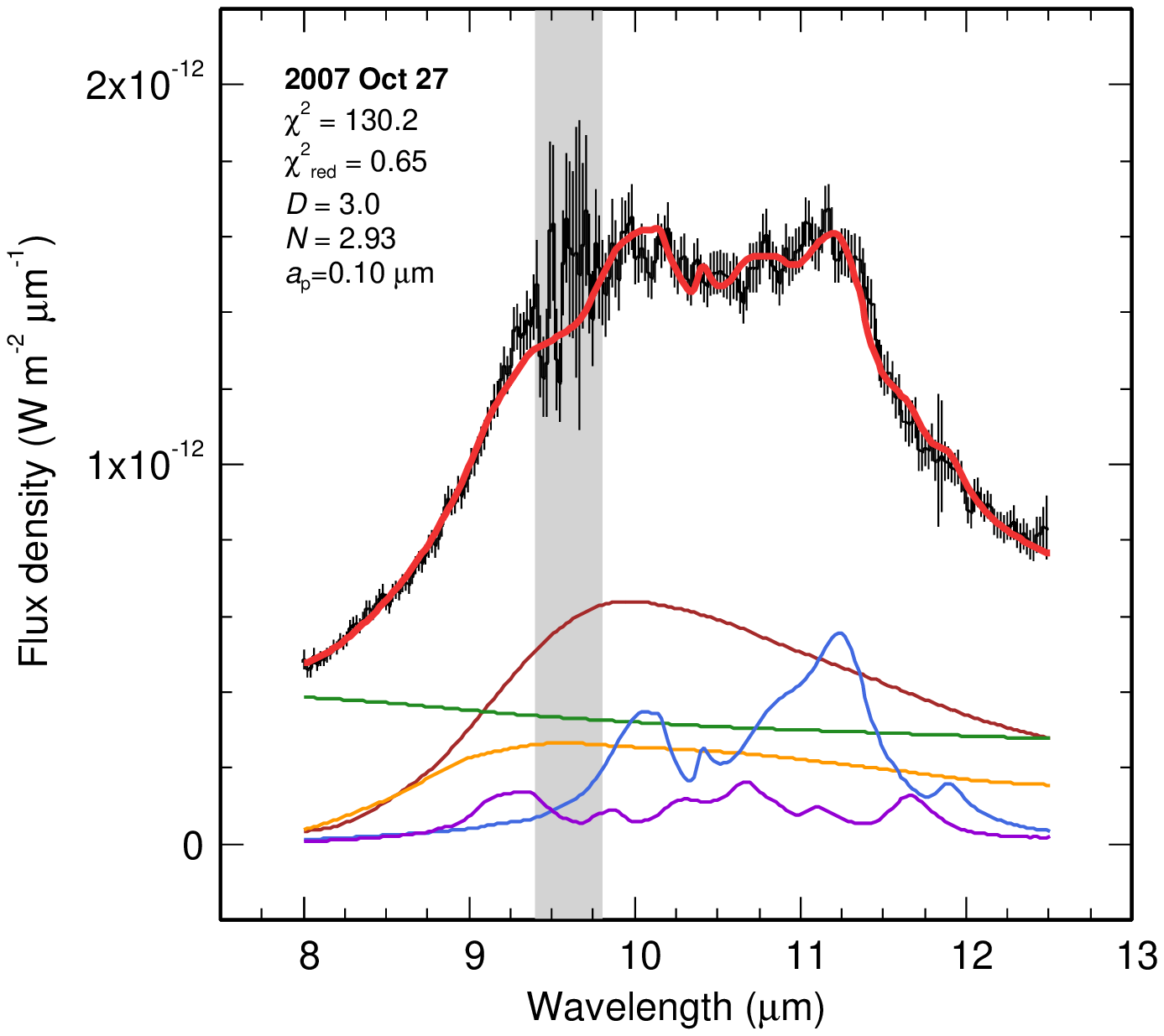}{0.45\textwidth}{(c)}
          \fig{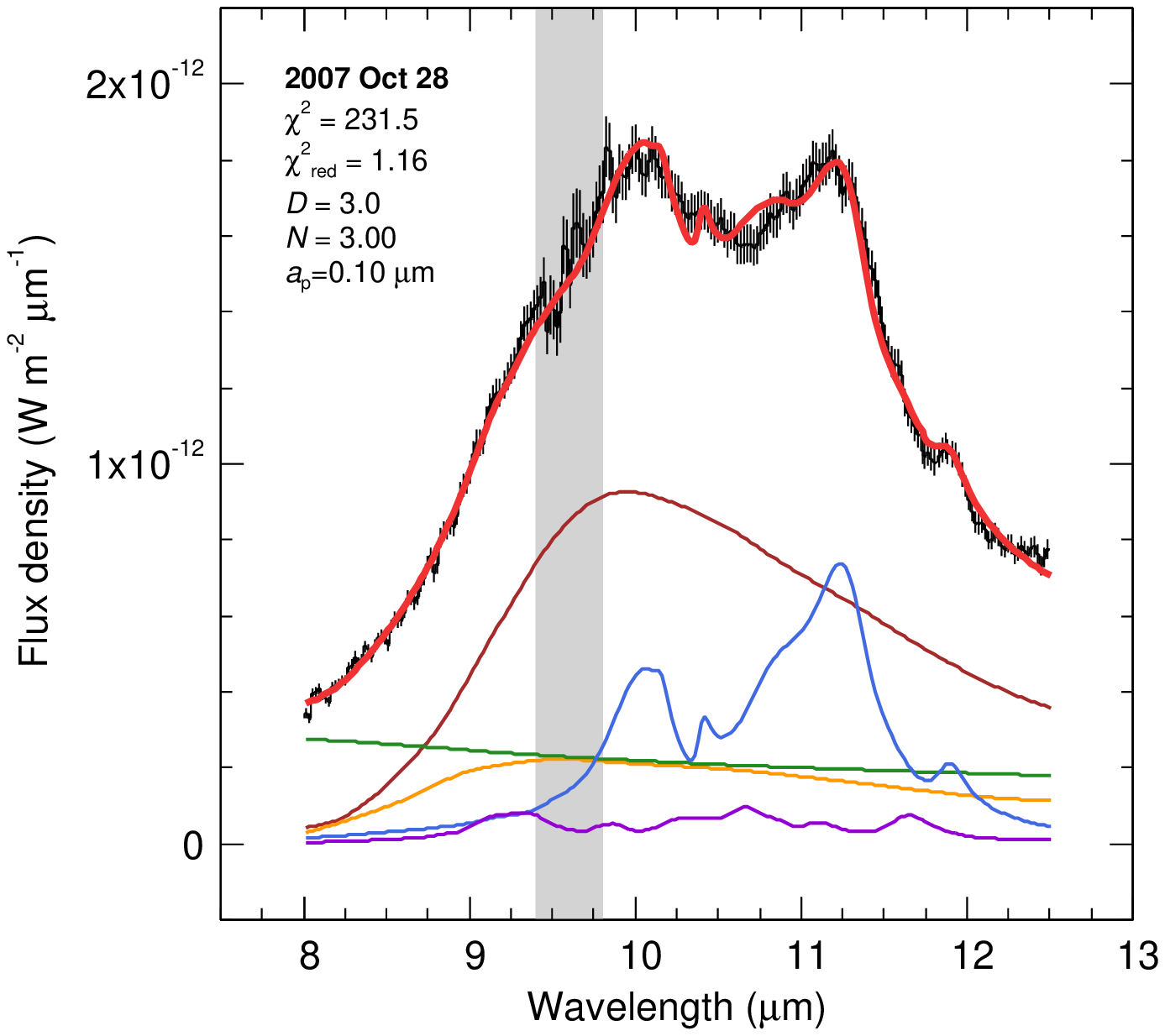}{0.45\textwidth}{(d)}
          }
\caption{Best-fit thermal models for comet 17P/Holmes from mid-infrared spectra on UT 2007 October 25 (a), 26 (b), 27 (c), and 28 (d). Black bars are observed spectra with error. Thick red line indicates best-fitted synthesized spectra reproduced by the thermal emission model. Thin brown, yellow, green, blue, and violet lines indicate best-fitted spectra of amorphous olivine, amorphous pyroxene, amorphous carbon, crystalline olivine and crystalline pyroxene, respectively. Gray hatches between 9.4 and 9.8 $\mu$m is absorption of telluric ozone (O$_{3}$). In left upper side of each panel, fitting results ($\chi^{2}$ and reduced-$\chi^{2}$) and best-fit dust properties (fractal dimension of dust grains, $D$, size distribution of dust grains, $N$, and peak radius, $a_{\rm p}$) are listed.\label{fig:fig1}}
\end{figure*}

\begin{figure}
\plotone{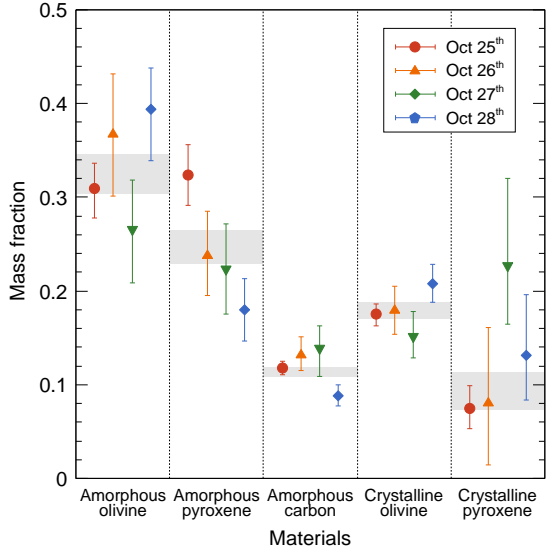}
\caption{Mass fraction of each dust material of comet 17P/Holmes on UT 2007 October 25 (circle), 26 (upper triangle), 27 (lower triangle) and 28 (diamond). Gray hatches of each material indicate weighted means and 1$\sigma$ error level of mean errors.  \label{fig:fig2}}
\end{figure}

\begin{figure}
\plotone{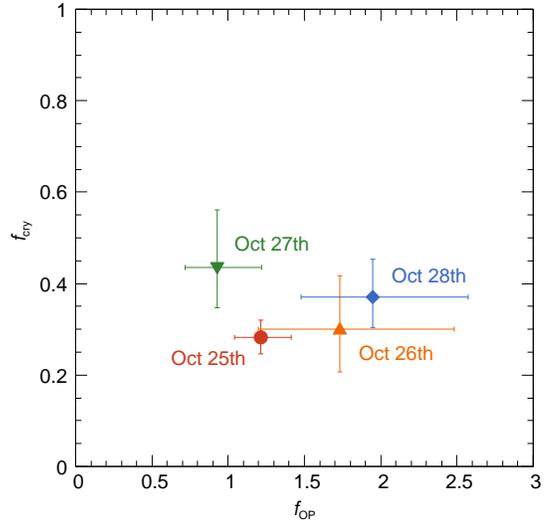}
\caption{Derived mass fractions of crystalline silicates, $f_{\rm cry}$, (vertical axis) and olivine-to-pyroxene abundance ratios, $f_{\rm OP}$, (horizontal axis) of comet 17P/Holmes on UT 2007 October 25 (circle), 26 (upper triangle), 27 (lower triangle) and 28 (diamond).  \label{fig:fig3}}
\end{figure}

Figure \ref{fig:fig1} shows the observed mid-infrared low-resolution spectra of comet 17P/Holmes on UT 2007 October 25, 26, 27, and 28. The 10-$\mu$m silicate feature is clearly recognized in all spectra. We apply the thermal emission model for cometary dust grains to the spectra. The best-fit results for the mass fractions of sub-micron grains for, dust properties, and minimum reduced-$\chi^{2}$ are listed in Table \ref{tab:table2} and the best-fit modeled spectra are shown in Figure \ref{fig:fig1}. Error ranges of each mass fraction are estimated as the 68.3\% confidence intervals based on the number of free parameters in the thermal emission model (i.e., $\Delta\chi^{2}$ = 9.31). Mass fractions of five minerals on each date are plotted in Figure \ref{fig:fig2} and listed in Table \ref{tab:table2}. We also derive a mass fraction of crystalline components in silicates ($f_{\rm cry}$) and an olivine-to-pyroxene abundance ratio ($f_{\rm OP}$) in comet 17P/Holmes. The $f_{\rm cry}$ and $f_{\rm OP}$ are given by
\begin{equation}
f_{\rm cry} = \frac{m_{\rm CryOl} + m_{\rm CryPy}}{m_{\rm AmoOl} + m_{\rm AmoPy} + m_{\rm CryOl} + m_{\rm CryPy}}
\end{equation}
and 
\begin{equation}
f_{\rm OP} = \frac{m_{\rm AmoOl} + m_{\rm CryOl}}{m_{\rm AmoPy} + m_{\rm CryPy}} , 
\end{equation}
respectively. The abundance of amorphous carbon is not considered to determine  $f_{\rm cry}$ and $f_{\rm OP}$. Derived  $f_{\rm cry}$ and $f_{\rm OP}$ for comet 17P/Holmes are also listed in Table \ref{tab:table3} and shown in Figure \ref{fig:fig3}. To compare with our results, we summarize dust properties and recalculated  $f_{\rm cry}$ and $f_{\rm OP}$ of comets reported in previous studies using a similar thermal emission model \citep{Ha02, Ha04, Ha05, Oo07a, Oo07b, Wo04, Wo11}.

\begin{figure}
\plotone{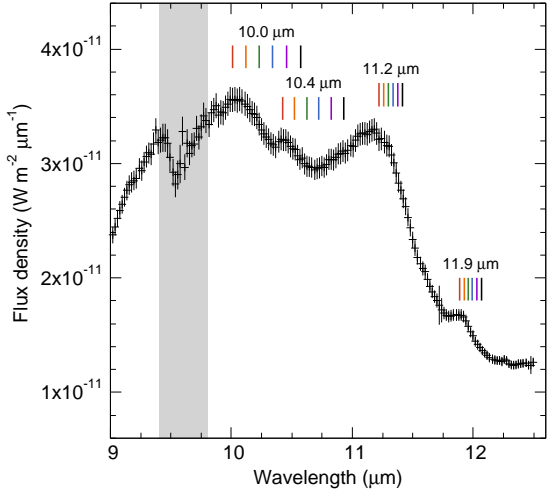}
\caption{Expanded spectrum of 17P/Holmes taken on 2007 October 25. The pairs of six vertical lines (red, orange, green, blue, purple, and black in the order left to right) indicate peak wavelength of 10.0 $\mu$m, 10.4 $\mu$m, 11.2 $\mu$m, and 11.9 $\mu$m with the Mg/(Mg+Fe) ratio of 100\%, 80\%, 60\%, 40\%, 20\%, and 0\%, respectively. Peak wavelengths of each Mg/(Mg+Fe) ratio are calculated from the expressions listed in Table 3 of \citet{Ko03}. Gray vertical hatch is the absorption band of telluric ozone. 
 \label{fig:fig4}}
\end{figure}

We assumed Mg-rich grains as crystalline silicate in our thermal emission model because cometary crystalline silicate grains are known to be Mg-rich (e.g., \citealt{Wo00}). The wavelengths of each sub-peak of the 10-$\mu$m feature for particles of crystalline olivine depend on Mg/(Mg + Fe) ratios \citep{Ko03}. We found that the assumption of the model, which Mg-rich grains as crystalline silicate, is correct based on the comparison of wavelengths of sub-peaks in the observed spectra with linear fitting equations between peak position and Mg/(Mg + Fe) ratios of crystalline olivine (Table 3 of \citealt{Ko03}) (see Figure \ref{fig:fig4}). 

Both the derived mass abundance fraction of crystalline in silicate, $f_{\rm cry}$, and the olivine-to-pyroxene abundance ratio, $f_{\rm OP}$, a few days after the outburst of comet 17P/Holmes shows no apparent temporal variation. Weighted means of $f_{\rm cry}$ and $f_{\rm OP}$ of the comet are 0.31 $\pm$ 0.03 and 1.20$^{+0.16}$/$_{-0.12}$, respectively. \citet{Re10} also reported the model-fit with their mid-infrared spectra (from 5.2 to 38 $\mu$m) taken on 10 November 2007 by the $Spitzer$ Space Telescope. The crystalline silicate features were also recognized in their spectrum. However, they did not use amorphous olivine for their model-fit, and we cannot, unfortunately, compare our results of model-fit with their results.

\section{Discussion} \label{sec:discussion}
\subsection{Formation region of dust of comet 17P/Holmes} \label{subsec:formation_region}

   As listed in Table \ref{tab:table3}, the dust grains of comet 17P/Holmes are compact ($D$ = 3.0), abundant in sub-$\mu$m-sized dust particles ($a_{\rm p} \sim$0.1 $\mu$m and $N \sim$3; a smaller peak size and a smaller power-law index of the size distribution for sub-$\mu$m sized dust grains) compared to other comets. Temporal evolution of size distributions for sub-micron-sized dust grains in comet 17P/Holmes (see Table \ref{tab:table2}) is not significant. It is likely that these properties of dust grains are almost intrinsic and pristine (not affected by the mega-burst events, such as fragmentation). Because a typical grain size increased with time by collisional growth in the solar nebula, such small dust grains in comet 17P/Holmes might form in an earlier phase or in the outer part of the solar nebula. In contrast with small grains (accessed by the 10-$\mu$m silicate feature), it is demonstrated that larger dust grains (mm-size and larger) have a steeper power-law index of size distribution \citep{Is10, Bo12}.

In comparison with other comets, comet 17P/Holmes shows smaller $f_{\rm cry}$ among comets found in previous articles (Table \ref{tab:table3}). In general, it is thought that the smaller $f_{\rm cry}$ of a comet indicates further distances from the Sun for the comet-formation (the larger $f_{\rm cry}$ for closer distances from the Sun). We conclude that comet 17P/Holmes formed in the outer region of the solar nebula. An enrichment of hyper-volatile species (such as CH$_{3}$OH, C$_{2}$H$_{6}$, C$_{2}$H$_{2}$) compared to other comets was demonstrated by near-infrared observations immediately after the outburst of the comet \citep{DR08, DR16} may support our hypothesis. Furthermore, some mechanisms of outburst of 17P/Homes were suggested \citep{Re10, Sc09, Boi02, Br04, de16}, and highly volatile abundant species (such as CO, CO$_{2}$) are more likely to contribute to the outburst of 17P/Holmes significantly.

\begin{deluxetable*}{ccccccccc}[ht!]
\tablenum{3}
\tablecaption{Fitting results of the thermal spectra of 17P/Holmes by using the thermal model\label{tab:table3}}
\tablewidth{0pt}
\tablehead{
\colhead{Comet}   & \colhead{UT Date} & \colhead{$r_{\rm H}$} & \colhead{$D$} & \colhead{$a_{\rm p}$ ($\mu$m)} & \colhead{$N$} & \colhead{$f_{\rm cry}$}        & \colhead{$f_{\rm OP}$} & \colhead{References}              
}
\startdata
17P/Holmes  & Weighted mean       & 2.45 & ---   & ---                       & ---                       & 0.31$\pm$0.03            & 1.20$^{+0.16}/_{-0.12}$  & This work \\
            & 2007 Oct 25         & 2.44 & 3.0   & 0.10$^{+0.01}/_{-0.00}$   & 3.34$^{+0.09}/_{-0.03}$   & 0.28$\pm$0.04            & 1.21$^{+0.20}/_{-0.17}$  & This work \\
            & 2007 Oct 26         & 2.45 & 3.0   & 0.10$^{+0.03}/_{-0.00}$   & 3.07$^{+0.14}/_{-0.03}$   & 0.30$^{+0.12}/_{-0.10}$  & 1.73$^{+0.75}/_{-0.53}$  & This work \\
            & 2007 Oct 27         & 2.45 & 3.0   & 0.10$^{+0.12}/_{-0.00}$   & 2.93$^{+0.44}/_{-0.03}$   & 0.44$^{+0.13}/_{-0.09}$  & 0.93$^{+0.29}/_{-0.21}$  & This work \\
            & 2007 Oct 28         & 2.45 & 3.0   & 0.10$^{+0.14}/_{-0.00}$   & 3.00$^{+0.18}/_{-0.02}$   & 0.37$^{+0.08}/_{-0.07}$  & 1.95$^{+0.63}/_{-0.47}$  & This work \\
9P/Tempel 1 & 2005 Jul 4 (1.0 h\tablenotemark{b}) & 1.51 & 2.857 & 0.3      & 3.7                       & 0.13\tablenotemark{c}    & 0.92\tablenotemark{c}    & 1 \\
            & 2005 Jul 4 (1.8 h\tablenotemark{b}) & 1.51 & 2.857 & 0.5      & 3.7                       & 0.36\tablenotemark{c}    & 7.22\tablenotemark{c}    & 1 \\
            & 2005 Jul 4 (3.5 h\tablenotemark{b}) & 1.51 & 2.857 & 0.4      & 3.6                       & 0.83$\pm$0.10            & 6.5$\pm$1.9              & 2 \\
73P-B/SW3\tablenotemark{a}        & 2006 Apr 29  & 1.11   & 2.727 & 0.5      & 3.4                       & 0.45$\pm$0.21            & 0.25$\pm$0.16            & 3 \\
73P-C/SW3\tablenotemark{a}        & 2006 Apr 30  & 1.09   & 2.727 & 0.3      & 3.4                       & 0.52$\pm$0.13            & $>$17                    & 3 \\
C/1995 O1   & 1996 Oct 11-14      & 2.8  & 2.8   & 0.2                       & 3.4                       & 0.53$\pm$0.04            & 2.65$\pm$0.51            & 4 \\
            & 1997 Feb 14-15      & 1.21 & 2.5   & 0.2                       & 3.7                       & 0.47$\pm$0.01            & 1.55$\pm$0.07            & 4 \\
            & 1997 Apr 11         & 0.97 & 2.5   & 0.2                       & 3.7                       & 0.62$\pm$0.02            & 2.26$\pm$0.17            & 4 \\            
            & 1997 Jun 24-25      & 1.7  & 2.7   & 0.2                       & 3.7                       & 0.56$\pm$0.04            & 1.57$\pm$0.08            & 4 \\
C/2001 Q4   & 2004 May 11         & 0.97 & 3.0   & 0.3                       & 3.7                       & 0.70\tablenotemark{c}    & 3.57\tablenotemark{c}    & 5 \\
            & 2004 Jun 4          & 1.02 & 3.0   & 0.2                       & 3.6                       & 0.71\tablenotemark{c}    & 6.88\tablenotemark{c}    & 6 \\
C/2002 V1   & 2003 Jun 10         & 1.18 & 2.857 & 0.5                       & 3.5                       & 0.66\tablenotemark{c}    & 2.63\tablenotemark{c}    & 6 \\
C/2007 N3   & 2009 Mar 3          & 1.45 & 2.73  & 0.9                       & 4.2                       & 0.43$\pm$0.15            & 0.35$\pm$0.11            & 7 \\
\enddata
\tablenotetext{\tiny a}{we refer to the results of the extraction boxes of opt-center (offset of 0 arcsec) for comet 73P/Schwassmann-Wachmann 3, corresponding to apertures B and M in Figure 2 of \citet{Ha11} for fragments B and C of the comet 73P/SW3, respectively.}
\tablenotetext{\tiny b}{time after the impact of the Deep Impact Mission in hours.}
\tablenotetext{\tiny c}{there is no description about error estimation in the applicable paper.}
\tablecomments{$f_{\rm cry}$ and $f_{\rm OP}$ of all comets are calculated by expressions (1) and (2), respectively.}
\tablenotetext{\tiny}{{\bf References.} [1] \citet{Ha05}, [2] \citet{Oo07b}, [3] \citet{Ha11} [4] \citet{Ha04} (Erratum of \citealt{Ha02}), [5] \citet{Wo04}, [6] \citet{Oo07a}, [7] \citet{Wo11}. }
\end{deluxetable*}

\begin{deluxetable}{lcc}
\tablenum{4}
\tablecaption{mass fraction of crystalline silicate, $f_{\rm cry}$, for each material of 17P/Holmes \label{tab:table4}}
\tablewidth{0pt}
\tablehead{
\colhead{UT Time}  & \multicolumn2c{$f_{\rm cry}$}  \\ 
\colhead{}         & \colhead{Olivine} & \colhead{Pyroxene}           
}
\startdata
2007 Oct 25 & 0.36 $\pm$ 0.02 & 0.19 $\pm$ 0.06 \\
2007 Oct 26 & 0.33 $\pm$ 0.03 & 0.24 $\pm$ 0.17 \\
2007 Oct 27 & 0.38 $\pm$ 0.02 & 0.56 $\pm$ 0.09 \\
2007 Oct 28 & 0.36 $\pm$ 0.01 & 0.46 $\pm$ 0.09 \\
Mean        & 0.36 $\pm$ 0.01 & 0.34 $\pm$ 0.04 \\
\enddata
\end{deluxetable}

Here we compared our fitting results with theoretical predictions by \citet{Ga01, Ga04} and \citet{Boc02}, considering equilibrium crystallization near the Sun as dust formation and mass transportation from the inner to outer region in the solar nebula. However, the fraction of crystalline silicate grains predicted by these models may not be reliable as absolute values although their trends (smaller $f_{\rm cry}$ for further distance from the Sun) are reasonable. The formation scenario of crystalline silicate grains in the early solar system and how to derive them into cometary nuclei are still in debate (e.g., \citealt{TaNo15}), and we strongly encourage comparison of further theoretical studies with cometary observations. \citet{Pi16} reported that the proto-planetary disk of HL Tau has small turbulent viscosity coefficient of a few 10$^{-4}$ by radio observations, implying that the turbulence might not be so large as to be able to transport a large amount of crystalline silicate grains to comet-forming region (further from the snowline of water) even if turbulence occurred in the solar nebula. Thus, it is in need of other transport mechanisms to the comet-forming region; e.g., X-wind, vertical mixing in the solar nebula \citep{Sh97}.

\subsection{Formation mechanisms of cometary dust} \label{subsec:formation_mechanisms}

   In the thermal emission model, we assume that mass ratio of olivine originating in interstellar medium (ISM) and solar nebula (SN) is equal to that of pyroxene. If this assumption is incorrect, there is no reason for $f_{\rm cry}$ of olivine to become the same as that of pyroxene in general. We may have to consider different transportation mechanisms for different materials. For instance, grain properties such as typical size and porosity might be different between olivine and pyroxene. In this case, the efficiency of transportation by radiation pressure for olivine is expected to be different from that for pyroxene \citep{TaNo15}. Table \ref{tab:table4} shows that values of $f_{\rm cry}$ for both olivine ($f_{\rm cry,olivine} = m_{\rm CryOl} / (m_{\rm AmoOl} + m_{\rm CryOl})$) and pyroxene ($f_{\rm cry,pyroxene} = m_{\rm CryPy} / (m_{\rm AmoPy} + m_{\rm CryPy})$) in all date are in agreement within 3$\sigma$ error bars. Based on this result, we consider that mass ratio of origin of olivine (ISM/SN mass ratio) is equal to that of pyroxene as working hypothesis to discuss the difference in $f_{\rm cry}$ between olivine and pyroxene.

We assume that dust grains originating in ISM and in the inner solar nebula ($<$0.1 au) have $f_{\rm cry}$ = 0.2\% $\pm$ 0.2\% and $f_{\rm OP}$ = 5.6 considering the observation toward Galactic center sources \citep{Ke04} and $f_{\rm cry}$ = 100\% and $f_{\rm OP}$ = 0.1 derived by theoretical calculation based on the equilibrium condensation condition \citep{Ga04}, respectively. We define two parameters, $x$ and $\gamma$, where the mass ratios of dust from the ISM or from the SN are $x$ (0 $\leq x \leq$ 1) and 1-$x$, respectively, and the mass ratios of dust from the ISM captured directly into the cometary nucleus or changing mineral form (e,g., crystallization and conversion from forsterite to enstatite) in the SN are $\gamma$ (0 $\leq \gamma \leq$ 1) and 1-$\gamma$, respectively.
Under these conditions, $f_{\rm cry}$ and $f_{\rm OP}$ of comets are given by
\begin{equation}
f_{\rm cry,comet} = \gamma xf_{\rm cry,ISM} + (1 - \gamma)xf_{\rm cry,SN} + (1-x)f_{\rm cry,SN}
\end{equation}
and 
\begin{equation}
f_{\rm OP,comet} = \gamma xf_{\rm OP,ISM} + (1 - \gamma)xf_{\rm OP,SN} + (1-x)f_{\rm OP,SN} , 
\end{equation}
where subscript X of $f_{\rm cry,X}$ and $f_{\rm OP,X}$ means comet as observed value in comet, ISM as interstellar medium, SN as solar nebula, respectively. The first term of the right-hand side of both formula is dust of ISM origin and was incorporated into the comet directly, the second term means dust of ISM origin that was denatured in SN, and the third term is dust originating in SN. We found that there is no solution that satisfies both expressions simultaneously, when we substitute the weighted means of $f_{\rm cry}$ and $f_{\rm OP}$ in 17P/Holmes (e.g., $f_{\rm cry}$ = 0.31 $\pm$ 0.03 and $f_{\rm OP}$ = 1.20$^{+0.16}$/$_{-0.12}$). For other comets for which both $f_{\rm cry}$ and $f_{\rm OP}$ were reported (summarized in Table \ref{tab:table3}), there is no unique solution. These results may support the interpretations that most dust grains of comets including 17P/Holmes did not form by the equilibrium condensation and annealing directly from gas-phase in the solar nebula and/or mass ratio of olivine originating in both ISM and SN are different from those of pyroxene.

\acknowledgments

This paper is based on data collected at the Subaru Telescope and obtained from the SMOKA, which is operated by the National Astronomical Observatory of Japan. The authors sincerely thank Dr. I. Sakon for acquiring the invaluable mid-infrared data of comet 17P/Holmes used in this study. This research was supported by a Grant-in-Aid for Japan Society for the Promotion of Science Fellows, 15J10864 (YS) and for Scientific Research (C), 17K05381 (TO). The authors also thank Ms. Yolande McLean for improving our English of this paper. The authors sincerely thank the anonymous reviewer for constructive comments and suggestions that have helped us to improve our manuscript.

\end{document}